\documentclass[12pt,twoside]{article}
\usepackage{fleqn,espcrc1,epsfig}
\begin{document}
\title{Hadrons in dense and hot matter~: implications of chiral symmetry
restoration}

\author{Guy Chanfray
\\
Institut de Physique Nucl\'eaire de Lyon, IN2P3-CNRS et Universit\'e
Claude-Bernard Lyon I, 43 Bd du 11 Novembre 1918, 69622 Villeurbanne Cedex,
France}
\maketitle
\vskip 1. true cm
\begin{abstract}
Recent developments on medium modifications of hadron properties in dense and hot 
hadronic matter are discussed. I will focus in particular on the behavior of
spectral functions associated to
collective scalar-isoscalar modes, kaons and vector mesons from ordinary nuclear
matter to highly excited matter produced in relativistic heavy-ion collisions from
SIS to SPS energies. Various theoretical approaches are presented in connection
with the interpretation of experimental data. Special  emphasis 
will be put on the role of
chiral dynamics and chiral symmetry restoration. I will discuss in
particular to which extent the broadening of the rho meson peak signals the onset
of chiral symmetry restoration.  
\end{abstract}
\section{Introduction}
The problem of the in-medium modifications of hadron properties have motivated
numerous works in the recent years both at the experimental   and  theoretical 
levels. One major goal of this rapidly developing field  is  the very
fundamental problem of progressive chiral symmetry restoration with increasing
temperature and/or baryonic density. In this talk I will try to draw some
conclusions on what has been established recently starting from very much discussed 
examples (rho and sigma mesons, kaons).
It will be also emphasized that relevant
information can be obtained from variety of sources ranging from intermediate
energy physics in the GeV range probing ordinary nuclear matter up to
ultrarelativistic heavy-ion collisions probing hadronic matter under extreme
conditions. 
 
\section{Chiral symmetry~: breaking and restoration}
Asymptotic freedom and color confinement are usually considered
as the most prominent properties of our theory of strong interaction,
Quantum Chromodynamics (QCD). However
QCD also  possesses an almost exact symmetry, the $SU(2)_L\otimes SU(2)_R$
chiral symmetry which is certainly the most important key for the
understanding of many phenomena in low energy hadron physics.
This symmetry originates from the fact
that the QCD Lagrangian is almost invariant under the separate flavor $SU(2)$
transformations  of
right-handed $q_R=(u_R,d_R)$ and left-handed $q_L=(u_L,d_L)$
light quark fields $u$ and $d$~:

\smallskip
\noindent
$q_R\to e^{i\vec\tau.\vec\alpha_R/2}\,q_R,\qquad 
q_L\to e^{i\vec\tau.\vec\alpha_L/2}\,q_L$.
 
\smallskip
\noindent
The small explicit violation of chiral symmetry
is given by the mass term
of the QCD Lagrangian which is ${\cal L}_{\chi SB}=-m_q\, (\bar u u+\bar d d)$, 
neglecting isospin violation. The averaged light quark mass $m_q=(m_u+m_d)/2\le$ 10 MeV,
the scale of explicit chiral symmetry breaking, has to be compared with typical
hadron masses of order $1$ GeV, indicating that the symmetry is excellent 
and in the exact chiral limit ($m_q=0$) left-handed and right-handed quarks
decouple. From the associated left-handed and right-handed conserved currents, 
one usually introduces two linear combinations, the vector and axial currents~:
\begin{equation}
{\cal V}^\mu_k=\bar q\,\gamma^\mu\, {\tau_k\over 2}\, q,\qquad 
{\cal A}^\mu_k=\bar q\,\gamma^\mu\,\gamma_5 \,{\tau_k\over 2}\, q\label{CURR}
\end{equation}
The corresponding charges $Q_k^V$ and $Q_k^A$ commute with the QCD hamiltonian. 
However at variance with the vector charges (which actually coincide with the
isospin operators) the axial charges of the QCD vacuum are not zero~:
$Q_k^A |0\rangle \ne 0$. Hence
the QCD vacuum does not possess the symmetry of the vacuum {\it i.e.} chiral
symmetry is spontaneously broken (SCSB). This key property of the QCD vacuum is
evidenced by a set of remarkable properties listed below. 
\begin{itemize}
\item{} The appearance of (nearly) massless goldstone bosons~: the pions with
extremely small mass compared to other hadrons.

\item{} The building-up of a chiral quark condensate~:
$\langle \bar q q \rangle =\langle \bar u u +\bar d d \rangle/ 2$
which explicitly mixes, in the broken vacuum, left-handed and right-handed
quarks ($\langle \bar q q \rangle =\langle \bar q_L q_R + \bar q_R q_L\rangle/2$).

\noindent
Another order parameter at the hadronic scale is the pion decay constant
$f_\pi=94$ MeV which is related to the quark condensate by the
Gell-Mann-Oakes-Renner relation~:
$-2 m_q <\bar q q>_{vac}=m^2_\pi f^2_\pi$ 
valid to leading order in the current quark mass. It leads to 
 a large negative value $<\bar q q>_{vac}\simeq$ -(240 MeV)$^3$ 
indicating strong dynamical breaking of chiral symmetry.
\item{} The absence of parity doublets. The normal Wigner realization of chiral
symmetry would imply a doubling of the hadron spectrum. Each hadron would have a
``chiral partner'' with opposite parity and (nearly) the same mass. This is
obviously not the case since the possible chiral partners (such as
$\pi(140)-\sigma(400-1200)$, $\rho(770)-a_1(1260)$, $N(940)-N^*(1535)$) 
show a large mass splitting $\Delta M=500$ MeV. 
\end{itemize}

\noindent
When hadronic matter is heated and compressed, initially confined quarks and 
gluons start to percolate between the hadrons to be finally liberated. This
picture is supported by lattice simulations showing that strongly interacting
matter exhibits a sudden change in energy- and entropy-density (possibly
constituting a true phase transition) within a narrow temperature window around 
$T_c=170$ MeV. This transition is accompanied by a sharp decrease of the quark
condensate indicating chiral symmetry restoration. However far before the
critical region, partial restoration should follow through the simple presence of
hadrons. Indeed, inside the hadrons the scalar density, originating either from
the valence quarks or from the pion scalar density (the virtual pion cloud),
is positive hence decreasing the quark condensate. Said differently, the
presence of hadrons locally restores chiral symmetry. This statement can be made
quantitative since, to leading order in hadron densities, the condensate evolves
according to \cite{DR90,CO92}:
\begin{equation}
R={\langle\langle \bar q q\rangle\rangle (\rho_h, T)\over
\langle \bar q q\rangle_{vac}}=1-\sum_h\,{\rho_h \Sigma_h\over f_\pi^2
m_\pi^2}.\label{DROP}
\end{equation}
Each hadron species present with scalar density $\rho_h$ contributes to the
dropping of the condensate through a characteristic quantity $\Sigma_h$ 
directly related to the integrated quark scalar density inside the hadron $h$~: 
$\Sigma_h/m=\int_h\, d{\bf r}\, \langle h| \bar u u+\bar d d |h\rangle$.

In nuclear matter the relevant quantity is the nucleon sigma commutator
$\Sigma_N\simeq 45$ MeV. Putting the numbers together  one finds
a $30 \%$ restoration at normal nuclear matter density.   The sigma commutator
and the dropping of the chiral condensate can be estimated with effective
theories in terms of the relevant degrees of freedom. It receives contribution
from valence quarks, scalar field and virtual pion cloud. There is  strong
indication (from model calculations and analysis of photon data) showing that the
major part of the nucleon sigma term comes from the pion cloud piece 
\cite{JA92,BI92,CH99}:
$ \Sigma_N^{(\pi)}=\int_N\, d{\bf r} \langle N| m^2_\pi\,\vec\Phi_\pi^2/2
|N\rangle\simeq 30$ MeV. 
The leading order formula (\ref{DROP}),  valid only for a non-interacting
medium can be promoted to an exact one by replacing in nuclear matter the
pion-nucleon sigma term by the full pion-nucleus sigma term per nucleon. 
Various higher order contributions have been examined. The role of short-range
correlations has been found to be weak \cite{DCE1} but pion exchange contribution
({\it i.e.} the modification of the pion scalar-density $\langle
\Phi^2_\pi\rangle$ itself related to the full longitudinal spin-isospin response and
p-wave collective pionic modes) yields an increase of the pion-nucleon sigma
term of about $5$ MeV at normal density \cite{MA00}. 
The  acceleration of chiral
symmetry restoration has not been found in the work of ref.\cite{LU00}
 based on an effective
chiral lagrangian. In the framework of relativistic theories short range repulsion 
(omega exchange) also yields a deceleration of symmetry restoration \cite{BR96}.

\par
Despite the fact that  the condensate is not an experimental observable,
it is hardly conceivable that such a strong modification of the QCD vacuum
should not have spectacular consequences on hadronic properties, namely on the
hadronic spectral functions. A number  of works have been devoted to the possible link
between the evolution of the masses and the condensates using various models
(sigma models , NJL model,..) most of the time at the mean field level. This
activity has culminated with the universal scaling laws proposed by Brown and rho, 
where hadron 
masses drop together with quark and gluon condensates \cite{SCAL}. However,  
the link between the evolution of the masses and the
condensates cannot be an absolute one. For instance the pionic piece of
the quark  condensate does not contribute  to the evolution of the mass
\cite{GU00}. It manifests
in a more subtle way through the mixing of the vector and axial-vector correlators. 
More generally the modification of hadronic spectral
functions is certainly not restricted to the shift of centroids of  mass
distributions. In that respect we will  discuss how chiral dynamics may generate a
softening/sharpening or a broadening of hadronic spectral functions with some
specific and highly debated examples (sigma, kaon, and rho mesons).
The key question is the relationship between the observed reshaping and chiral
symmetry restoration. One possible strategy to obtain this crucial connection 
is to make a simultaneous study of the spectral functions associated with chiral
partners. A very important example is the rho meson and the axial-vector
meson  $a_1$ and we will see that there is a mixing of the associated current
correlators trough the presence of the pion scalar density as already mentioned
just above. 

\section{Scalar-isoscalar modes in nuclear matter}  

There are at least two excellent reasons to study the in-medium modifications
of the pion-pion interaction in the scalar-isoscalar channel both being 
related to fundamental questions in present-day nuclear physics. 
The first one relies on the binding energy  of nuclear matter since a 
modification of the correlated two-pion exchange may have
some deep consequences on the saturation mechanism \cite{MA99}. The second one 
is the direct connection with chiral  symmetry
restoration. Such a restoration implies that   there must be a
softening of a collective scalar-isoscalar mode, usually called the sigma
meson, which becomes degenerate with its chiral partner {\it i.e.} the pion 
at  full restoration density, even if this meson is not well identified in the
vacuum. This also implies that at some density 
the sigma meson spectral function should exhibit a spectacular enhancement 
near  the two-pion threshold. This effect can be seen as a precursor effect
of chiral symmetry restoration  associated with large fluctuations 
of the quark condensate near phase transition \cite{HA99}.

\noindent
\begin{figure}[!ht]
    \centering\epsfig{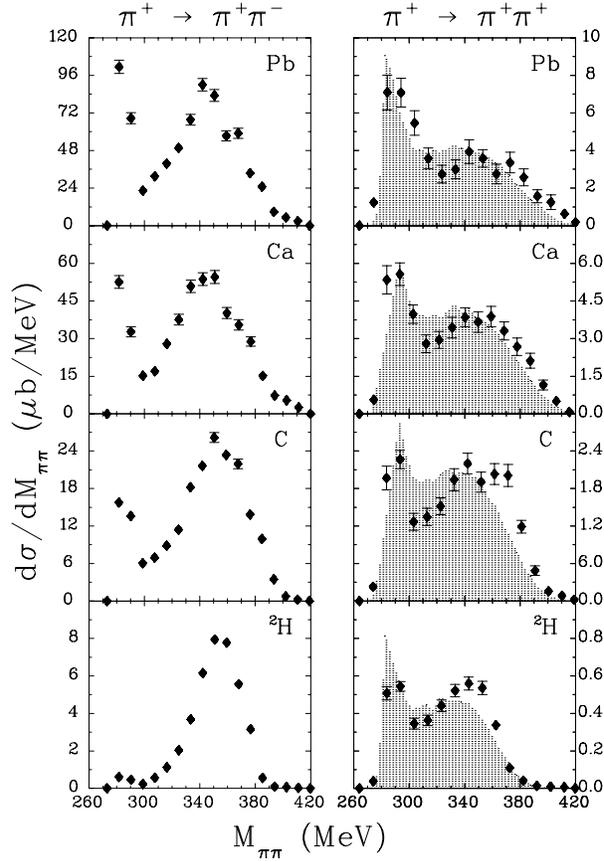}
  \caption{Invariant mass spectrum for the $\pi^+\to \pi^+\pi^-$ and
 $\pi^+\to \pi^+\pi^+$  mass distributions on various nuclei measured by the CHAOS
 collaboration \cite{BO00}. Diagrams (dots) are
 the results of phase-space simulations for the pion production in $\pi A\to 
 \pi\pi[A-1]$ reactions.} 
\end{figure}
\begin{figure}[!ht]
    \centering\epsfig{figure=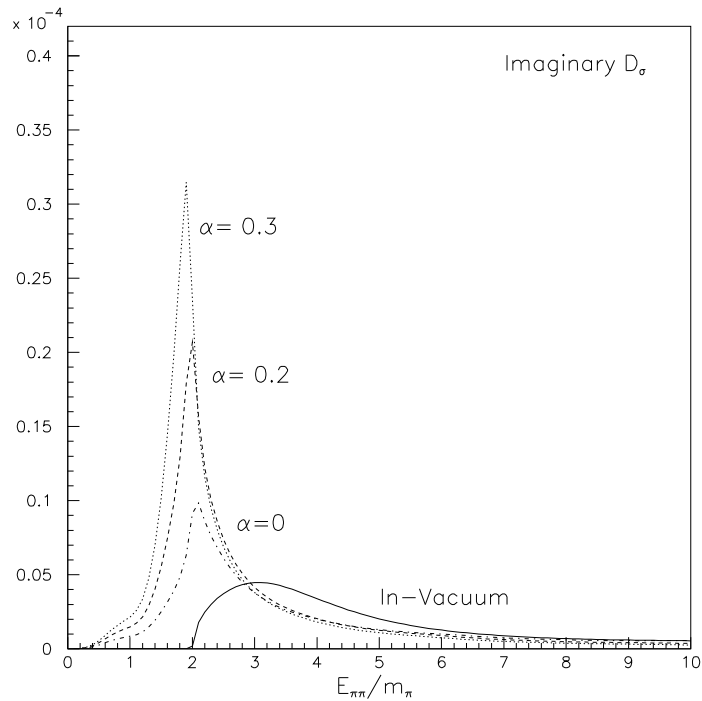,width=9.5cm}
  \caption{Spectral function of the sigma meson in vacuum (full curve) and at
  normal nuclear matter density (the three other curves). The dashed curve
  ($\alpha=0$) includes only p-wave effects and the other ones ($\alpha=0.2, 0.3$)
  incorporate a dropping sigma mass as explained in the text.} 
 \end{figure}
 
 \begin{figure}[!ht]
    \centering\epsfig{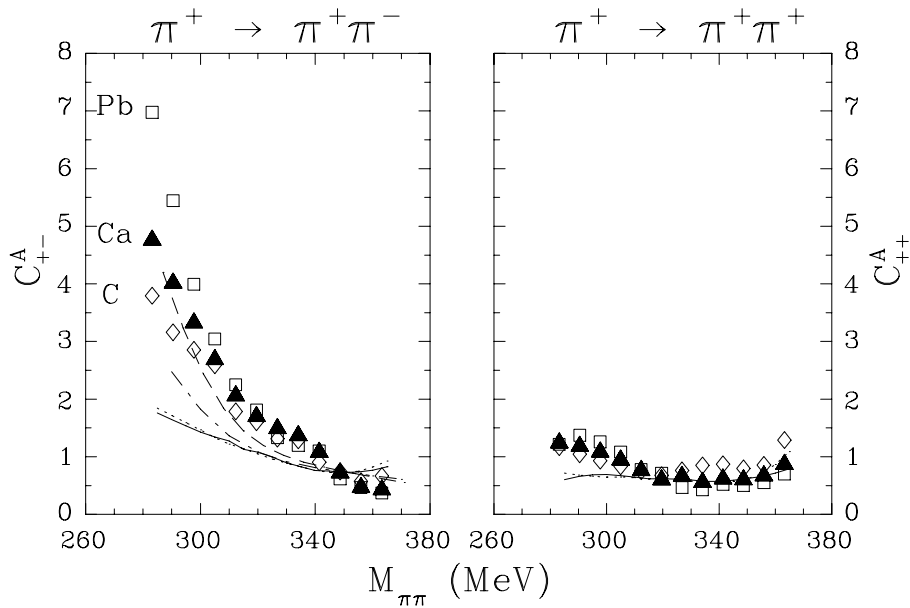}     
  \caption{The composite ratio $C^A_{\pi\pi}$ for $^{12}$C, $^{40}$Ca and
  $^{208}$Pb compared with theoretical calculations including p-wave pionic effects
  only (full curve), chiral symmetry restoration only (dash-dotted) and both
  effects (dashed). Taken from ref.\cite{BO00}.}    
\end{figure}

The  medium effect  which has been first proposed is a consequence of 
the modification  of the two-pion 
propagator and unitarized $\pi\pi$ interaction from the
softening of the pion dispersion relation by p-wave coupling 
to $p-h$ and $\Delta-h$ states \cite{SC88}.
The existence of the collective pionic modes  produces a strong accumulation of 
strength near the two-pion threshold 
in the scalar-isoscalar channel around $\rho_0$. On the experimental side the CHAOS
collaboration has measured at TRIUMF the invariant mass distribution of the 
produced pion pair in  $A (\pi^+,\pi^+\pi^-)$ and 
$A (\pi^+,\pi^+\pi^+)$ for incoming
pions of energy $283$ MeV (fig.1) \cite{BO00}. In the $\pi^+ \pi^-$ channel a strongly A-dependent
accumulation of strength is growing up from hydrogen  to lead. 
This effect is
not present in the  $\pi^+ \pi^+$ channel which is is purely isospin $I=2$ 
contrary to the $\pi^+ \pi^-$ channel predominantly isoscalar. Since the angular
distribution shows that the $\pi^+\pi^-$ pair is almost in a pure s-wave state,
this process probes scalar-isoscalar modes inside the nucleus. 
According to the first realistic  calculations \cite{SC98}, this reshaping of the strength 
coming from the p-wave pionic collective modes  may provide  a partial 
explanation  of the CHAOS data. These last results
have been  questioned in another recent paper  where  it is found
that pion absorption forces the reaction to occur at   very peripheral
density \cite{VI99}. Hence the effect of chiral
symmetry restoration has to be included on top of p-wave pionic effects 
to reach a better description of the data. This has been achieved in recent works
using the linear sigma model  implemented by adding a one-parameter 
form factor $v(k)$ to fit the phase shifts in vacuum once the scattering 
amplitude is unitarized \cite{ZH00}.  The (in-medium) unitarized scalar-isoscalar 
$\pi\pi$ T matrix (in the CM frame   and at total  energy $E$
of the pion pair) has been taken as \cite{SC98,ZH00}~:     
\begin{equation}
\langle{\bf k}, -{\bf k}|T(E)|{\bf k}', -{\bf k}'\rangle=v(k) v(k')\,
{6 \lambda (E^2-m^2_\pi)\over 1-3\lambda \Sigma(E)}
\,\bigg( E^2-m^2_\sigma- {6 \lambda^2 f^2_\pi \Sigma(E)\over 
1-3\lambda \Sigma(E)}\bigg)^{-1}\label{TMAT}
\end{equation}
where $\lambda$ is the $\sigma\pi\pi$ coupling and the last factor in (\ref{TMAT}) 
is nothing but the unitarized sigma meson propagator $D_\sigma(E)$ 
({\it i.e} with two-pion loop). 
The p-wave collective effects are embedded in the  two-pion loop~:
\begin{equation}
\Sigma(E)=\int {d{\bf q}\over (2\pi)^3}\, v(q)\,
\int{i \,dq_0\over 2\pi}\, D_\pi({\bf q}, q_0)\, D_\pi(-{\bf q},\, E-q_0).
\end{equation}
The pion propagator $D_\pi({\bf q}, q_0)$ is calculated in a standard 
nuclear matter approach  and incorporates the p-wave coupling of the pion to
delta-hole states with short-range screening described by the usual
$g'_{\Delta\Delta}=0.5$ parameter. Chiral symmetry restoration can be accounted for 
by dropping the sigma mass according to~:
$m_\sigma(\rho)=m_\sigma\,(1-\alpha \,\rho/\rho_0)$.
Such a density dependence very 
naturally arises in this model 
from the tad-pole graph where the sigma meson directly couples to the nuclear density. 
In \cite{HA99} a value of $\alpha$ in the range of 0.2 to 0.3 was found. 
The result for the invariant-mass distribution $Im D_{\sigma}(E_{\pi\pi})$ is shown 
in fig.2 at saturation density. One observes a dramatic downward shift of the mass 
distribution as compared to the vacuum. The low-energy enhancement,
already present without sigma-mass modification ($\alpha=0$) and induced by the 
density dependence of the pion loop,
is strongly reinforced as the in-medium $\sigma$-meson mass is included. 
For $\alpha=0.2$ and  $\alpha=0.3$ the peak height is increased by a factor 2 and 4 
respectively. Similarly for the T-matrix, a sizable effect can be noticed in its 
imaginary part which might be sufficient 
to explain the findings of the CHAOS collaboration. To facilitate the comparison with
theoretical calculations  the experimental group has  presented his data
in the form of a composite ratio \cite{BO00}:
\begin{equation}
C^A_{\pi\pi}=\big(M^A_{\pi\pi}/\sigma^A_T\big)\,/\, \big(M^N_{\pi\pi}/\sigma^N_T\big)
\end{equation} 
where $\sigma^A_T$ ($\sigma^N_T$) is the measured total cross section of the
$(\pi, 2 \pi)$ process in nuclei (nucleon). As  is apparent on fig.3, while a
conventional calculation is able to reproduce the $\pi^+\pi^+$ case, chiral
symmetry restoration on top of p-wave effects is needed to reach a better
agreement with data in the $\pi^+\pi^-$ case. 
It is therefore very tempting to conclude that a precursor effect of chiral
symmetry restoration has been seen in this experiment. However one has to keep
in mind that this comparison with theoretical predictions is at best semi-quantitative
and a full calculation for the absolute yields incorporating all the complicated
reaction dynamics together with the relevant medium effects remains to be
done before  firm conclusions can be drawn. Nevertheless experimental efforts in that
direction has certainly to be encouraged since the still poorly known in-medium 
scalar field is of utmost importance in present-day nuclear physics.

\section{Kaons in dense matter}

The strong interest for studying kaon production in relativistic heavy-ion
collisions originates from two complementary reasons. The first one is the 
sensitivity of kaon properties and propagation to the state of the produced
matter and the second one is the fact that the basic kaon-nucleon interaction 
is governed to a large extent by the $SU(3)$ extension of chiral symmetry. Hence
kaon production provides a unique opportunity to study chiral dynamics in a dense
and hot medium. Recent data taken
in particular at SIS by the KaoS and FOPI collaborations have demonstrated 
sizable medium effects; the details and experimental data can be found in the
contribution of P.Senger in this present volume \cite{SE00}. 
The absence of in-plan flow and the quite strong 
azimuthal emission of kaons ($K^+$) together with the unexpected high 
sub-threshold $K^-/K^+$ ratio favorize a strong in-medium attraction for the 
anti-kaons ($K^-$) and a more moderate but 
 significant repulsion for the
kaon ($K^+$). This effect has been predicted by many theoretical calculations
based on chiral symmetric frameworks. From a chiral symmetry lagrangian taken at the
mean-field level, one can easily establish the leading order terms 
modifying kaon and anti-kaon masses at finite proton and neutron densities~:
\begin{eqnarray}
\Delta m_{K^+}&=&+{1\over 2}\,{1\over 4 f^2_\pi}
\bigg(3\,(\rho_p+\rho_n)+(\rho_p-\rho_n)\bigg)
-\,{\Sigma_{KN}\over 2 m_K\, f^2_\pi}\,\rho \label{NASS}\\
\Delta m_{K^-}&=&-{1\over 2}\,{1\over 4 f^2_\pi}
\bigg(3\,(\rho_p+\rho_n)+(\rho_p-\rho_n)\bigg)
-\,{\Sigma_{KN}\over 2 m_K\, f^2_\pi}\,\rho\,.\label{MASS}
\end{eqnarray}
The first contribution arises from vector current interaction between the
pseudo-scalar mesons and the nucleons and is  determined by chiral
symmetry alone (Weinberg-Tomozawa theorem). It gives the bulk of repulsion 
for the kaon and attraction for the anti-kaons. The second term of scalar 
nature is proportional to the not very well-known kaon-nucleon 
sigma term $\Sigma_{KN}$ and
yields attraction in both cases; it is actually to next order in the chiral
perturbation expansion and could be balanced by other repulsive terms at this order
but without altering too much the strong mass splitting of $S=1$ and $S=-1$ modes
in matter.
\noindent
\begin{figure}
  \begin{minipage}[b]{.60\linewidth}
    \centering\epsfig{figure=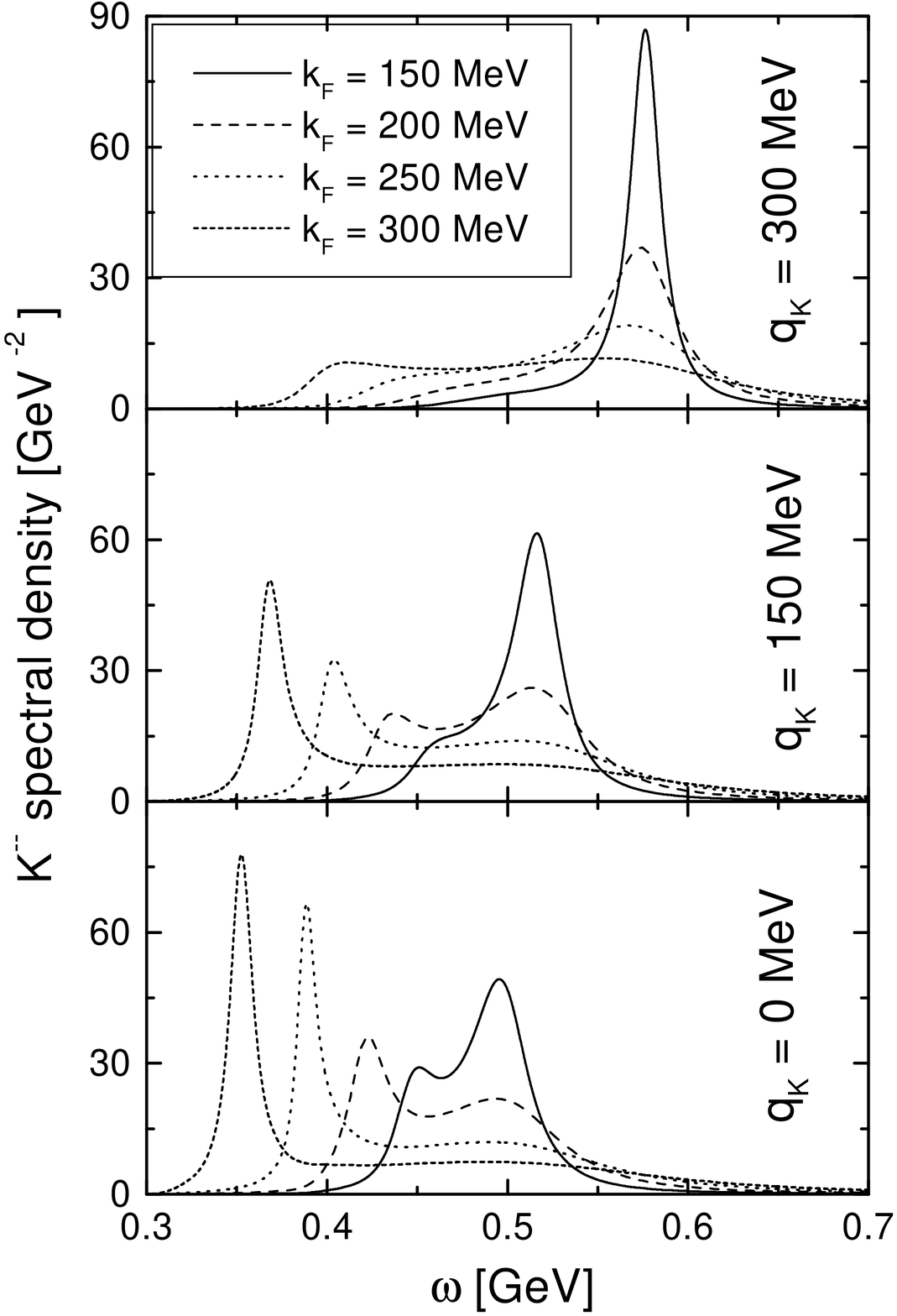,width=\linewidth}
  \end{minipage}\hfill  
  \begin{minipage}[b]{.70\linewidth}
    \centering\epsfig{figure=oset1.epsi,width=\linewidth}\end{minipage}  
\caption{{$K^-$ spectral function as a function of energy for several densities.
Left panel~: influence of the $K^-$ momentum in the self-consistent calculation
of ref. \cite{LUKA}. 
 Right panel~: zero momentum calculation of ref.\cite{RA00} with the influence of
in-medium dressing of the pion (c) on top of Pauli blocking (a) and in-medium
dressed kaons (b).}}
\end{figure} 

For the specific case of $K^+$, there is a consensus between the various
transport code approaches in favor of repulsion to account for the flow
variables data \cite{LI96,WA00,CA99}.  For what concerns the $K^+$ spectra and 
excitation functions 
in the sub-threshold or near-threshold region there is also a consensus for the
dominant role played by secondary processes \cite{CA99,FU97,AI00} 
such as $\pi N\to YK$ or $N\Delta\to
NYK$ ($Y=\Lambda, \Sigma$) which are very sensitive to modifications of in-medium
masses. Although, there are still uncertainties for the input cross-sections 
(this is especially important for the $N\Delta$ channel), transport code calculations
also favor in-medium repulsion for the $K^+$ \cite{FU99,AI00}. The sensitivity to
the EOS has also been  investigated. It turns out that  
when in-medium kaon mass is incorporated the sensitivity 
to the EOS decreases. In the case of soft EOS , the system reaches a higher
density yielding a higher $K^+$ production; this effect can be
counterbalanced by the larger increase of the $K^+$ mass increasing the threshold
production. Nevertheless, according to \cite{FA00} detailed study of Au-Au versus C-C
excitation functions with in-medium repulsion favors a soft EOS with 
compressibility of the of the order of   $200$ MeV.

Transport calculations \cite{LI97,SI97,CA99} 
also show that sub-threshold $K^-$ spectra are
better reproduced if a dropping  kaon mass of the type given by the mean field
approach (\ref{NASS},\ref{MASS}) is incorporated. It has been realized that the situation is not so
simple because the anti-kaon interaction in the medium is governed by a very rich and rather
complex chiral dynamics as we will discuss below. To begin with, we have to understand 
a rather paradoxal situation. On one hand, although the basic $K^- N$ Weinberg-Tomozawa is
attractive, the scattering lengths imply a repulsive interaction in vacuum. On the
other hand, kaonic atom data are compatible with a strongly attractive anti-kaon optical 
potential as large as 200 MeV once extrapolated at normal nuclear matter density. 
The solution of this paradox  
is the existence of a very peculiar object, the  $\Lambda(1405)$ resonance. 
Following ref.  \cite{WA96,LUKA,KOCH,RA00}
 one can start with a coupled channel equation for the vacuum 
$\bar K N$ scattering matrix, schematically written as~:
\begin{equation}
\langle \bar K N|T|\bar K N\rangle=\langle \bar K N|V_{WT}|\bar K N\rangle
+\langle \bar K N|V_{WT}|M B\rangle\, G_{M B}\,
\langle M B|T|\bar K N\rangle 
\end{equation}  
where $G_{M B}$, with $M(\bar K, \pi)$ and $B(N, \Lambda, \Sigma)$, 
is the meson-baryon propagator in the intermediate state. It turns out that 
the attractive  Weinberg-Tomozawa interaction ($V_{WT}$) in the isospin $I=0$ 
channel is sufficiently strong that it generates a pole at 27 MeV below the
$K^- p$ threshold. This pole correspond to a quasi-bound state, the
$\Lambda(1405)$ which decays into $\pi \Sigma$ with a width of about 50 MeV.
This is precisely the existence of this quasi-bound state what makes the
$K^-p$ interaction repulsive at threshold, while the Weinberg-Tomozawa
amplitude is attractive. Going at finite density  Pauli blocking starts to work
in the intermediate
$\bar K N$ state and  the quasi-bound state moves up above threshold. Hence the 
$\Lambda(1405)$ dissolves already at very small density \cite{WA96} making the in-medium 
$K^- N$ interaction attractive as seen in kaonic atom data.   However, as
pointed out by M. Lutz \cite{LUKA}, a self-consistent incorporation of the in-medium
dressed $K^-$ propagator in the intermediate state  modifies the picture. 
The position of the resonance does not really move up but it considerably
broadens. The in-medium $K^-$ spectral function exhibits a two-level structure,
the lower mode corresponding to the $K^-$ pole branch and the upper to the
$\Lambda(1405)$-hole branch with strength decreasing very fast with increasing
density (fig.4). Self-consistency makes the two-peak mode barely visible  and if
 the modification of the pion dispersion relation  is also included 
 the $K^-$ completely melts with 
the $\Lambda(1405)-h$ branch  even losing its status of quasi-particle \cite{RA00}.
Finally it has been emphasized that the observed enhancement of $K^-$ production 
has probably little relation with the dropping of the anti-kaon mass or in other
words to its optical potential at zero momentum \cite{KOCH}. Indeed, in the conditions
prevealing in heavy-ion collisions, the anti-kaons have a typical momentum of
$300$ MeV with respect to the matter rest frame. In that regime
a self-consistent calculation   shows little attraction if not repulsion
and the medium effect might originate   from the enhancement of 
cross-section of important
secondary processes such as $\pi \Sigma \to K^- p$ \cite{KOCH}. Although, there is not 
yet a quantitative
understanding of the $K^-$ production, the physics of the $K^-$ 
is a beautiful example of the 
very rich in-medium chiral dynamics. It is also an example of the strong reshaping 
and broadening of an hadronic spectral function which is also present but 
for different reasons  in the case of the rho meson.
 
\section{Dilepton production and the rho meson}

Low mass dilepton production has been reported as being among the evidences 
for the formation of a new phase of matter in relativistic heavy-ion 
collisions at CERN/SPS \cite{HE00}. In particular the CERES 
collaboration \cite{CERES} has observed an important radiation in the invariant mass region 
$300 -700$ MeV/c beyond what is expected from the conventional 
sources able to explain the proton-nucleus data (fig.5). Since these conventional 
sources (the so-called hadronic cocktail) correspond to final state Dalitz 
decays ($\eta, \eta'\to \gamma e^+ e^-$, $\omega\to \pi^0 e^+ e^-$) and 
direct vector meson decays ($\rho, \omega, \Phi\to e^+ e^-$), one can conclude 
that this excess of radiation originates from the interacting fireball before 
freeze-out. Due to the very large number of  produced pions the first 
candidate is the $\pi^+ \pi^-\to l \bar l$ annihilation process which is
dynamically enhanced by the rho meson. Using  vacuum meson  
properties many theoretical groups have included this process
within (very)
different models for the space-time evolution of $A-A$ reactions.
Their results are in reasonable agreement with each other, but in disagreement 
with the data~: the experimental spectra in the mass region $300-600$ MeV/c
are significantly underestimated and the rho peak itself has the tendency to be
overestimated as seen from fig.6. Thus one came to the conclusion that strong
medium effects yielding a flattening of the spectra are needed. This has motivated
a considerable theoretical activity that I will now briefly describe.
\begin{figure}[!ht]
    \centering\epsfig{figure=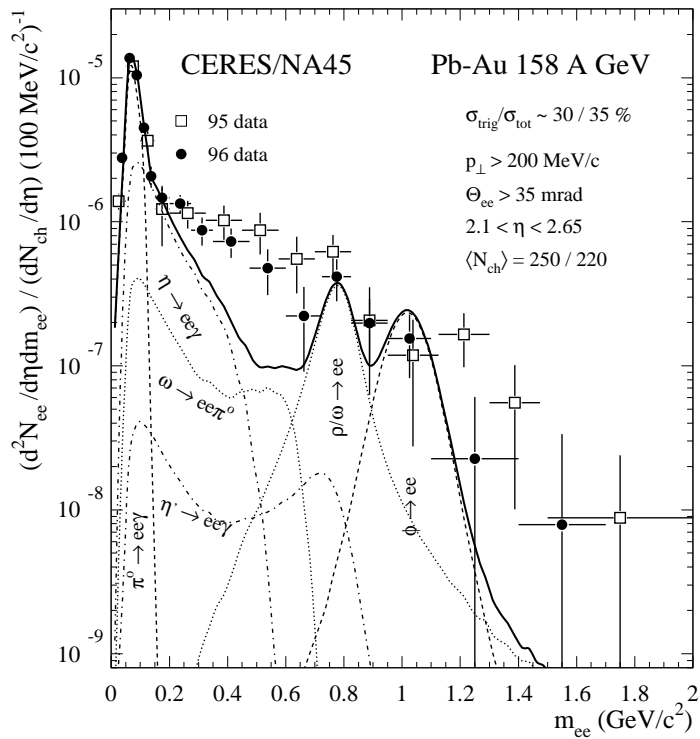,width=10.0cm}
  \caption{CERES/Na45 data on dilepton production data in central 158 AGeV
  Pb+Au  
  collisions compared  to a hadronic cocktail inferred from a thermal model.}  
\end{figure}  
\begin{figure}[!ht] 
    \centering\epsfig{figure=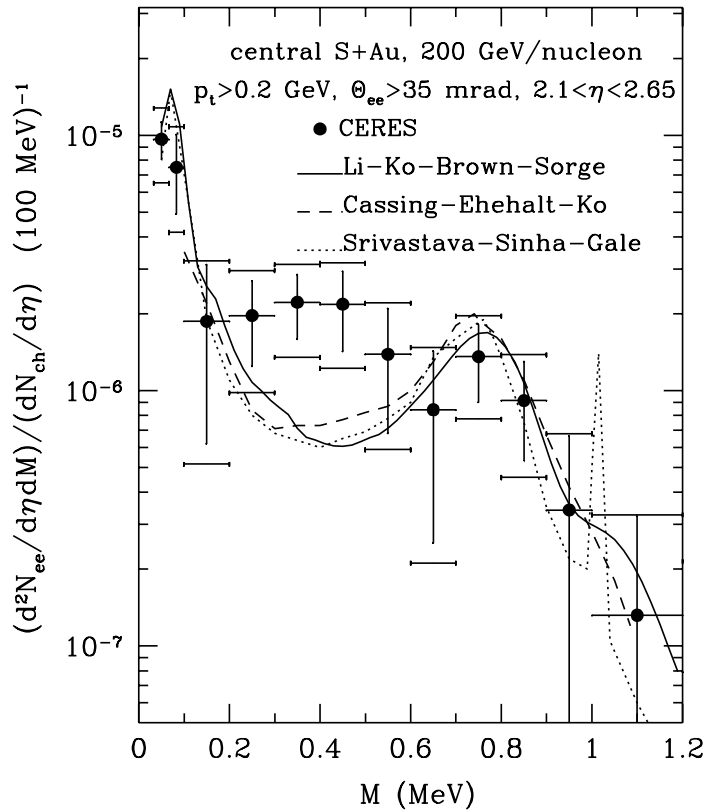,width=10.0cm}
  \caption{Dilepton invariant mass spectrum measured in central 200 AGeV S+Au
  collisions compared with theoretical calculations incorporating $\pi\pi$
  annihilitation with free space meson properties.}     
\end{figure}
\medskip
\noindent{\it Dilepton production from hot and dense matter}. The dilepton
production rate (DPR) 
per unit  4-volume from a hot ($T=1/\beta$) and dense medium is given by~:
\begin{equation}
{dN_{l\bar l}\over d^4x d^4 q }=-{\alpha^2\over 6 \pi^3 M^2}\,
{1\over e^{\beta q^0}-1}\,g_{\mu\nu}\,\left(-{1\over \pi}\,Im \Pi_V^{\mu\nu}\right) 
\end{equation}
where $M$ ($M^2=q^2_0-{\bf q}^2$) is the invariant mass of the produced pair.
Once the overall thermal factor has been extracted, the DPR is directly
proportional to the imaginary part of the current-current correlation
function~:
\begin{equation}
\Pi_V^{\mu\nu} (q)=-i\int\,d^4x\,e^{-iqx}\,\langle\langle J^\mu(x),J^\nu(0)
\rangle\rangle(T,\rho_B).
\end{equation}
For simplicity, we will concentrate on the (prevailing) isospin $I=1$
(isovector) projection of the electromagnetic current~:
\begin{equation}
{\cal V}^\mu_\rho={1\over 2}\,\left(\bar u \gamma^\mu u- \bar d \gamma^\mu d\right)
\end{equation}
which just coincides with the third component of the conserved vector current
of chiral symmetry (see eq.\ref{CURR}).  We know from the well established Vector Dominance phenomenology (VDM)
that the corresponding correlator is accurately saturated by the rho meson. 
This property is formally incorporated through the famous field-current identity
${\cal V}^\mu_\rho=(m^2_\rho/ g_\rho)\, \rho^\mu$.
Hence dilepton production allows to reach the imaginary part of the rho meson
propagator, namely the in-medium rho meson spectral function. To have some
insight about manifestation of chiral symmetry restoration this vector
correlator should be studied simultaneously with the axial-vector correlator in
which the properties of the chiral partner of the rho meson, namely the $a_1$
meson, are encoded. In the vacuum the SCSB manifests itself in the marked 
difference between the $\rho$ and the $a_1$ and the transition from the hadronic
to the partonic regime (``duality threshold'') is characterized by the onset of
perturbative QCD around $M_{dual}\simeq 1.5$ GeV. In the medium, full chiral symmetry
restoration requires the degeneracy of vector and axial correlators over the
entire mass range.

\medskip
\noindent
{\it Density expansion.} Several approaches have been put forward to determine
the spectral properties of vector mesons in the medium. One  method, very usual  in
nuclear physics, is the low density expansion~:
\begin{equation}
\Pi^{\mu\nu}(q, T, \mu)=\Pi^{\mu\nu}_{vac}(q)\,+\,
\sum_h \rho_h\,\Pi^{\mu\nu}_h(q).
\label{VIRIEL}
\end{equation}
Taking the imaginary part, one obtains the contribution of hadron species $h$
present with density $\rho_h$ to the spectral function. 
The vacuum piece is extremely well known from $e^+ e^-$ annihilation. The
hadronic part is expected to be dominated by the lightest meson ($\pi$) and baryon
($N$).  In the chiral reduction formalism \cite{ZA96}, the hadronic matrix elements
($\Pi^{\mu\nu}_h$) can be inferred from a combination of empirical information 
($\pi N$, $\rho N$ or $\gamma N$ data...) and chiral Ward identities. 
Although 
model independent in spirit, this framework  does not allow to perform
systematic resummations. Indeed it has the tendency to overestimate the rho
meson peak itself because these higher order many-body effects are absent. 
Nevertheless this approach explicitly contains the already mentioned axial-vector
mixing that we will now discuss.

\medskip 
\noindent{\it Axial-Vector mixing.} In the medium the emission and the
absorption of  thermal (finite temperature) or  virtual (finite density) pions 
is  able to transform a vector current into an axial current. In other words, 
the response of the system to a vectorial probe contains an axial contamination 
mediated by the pions, the pure vector piece being quenched by the emission and
absorption at the same point. Hence increasing temperature or density 
({\it i.e.} increasing the pion scalar density) makes the axial-vector mixing
more and more important until full restoration where axial and vector correlators
become identical. This mixing has been formally proven at finite temperature 
in the chiral limit,
using only chiral symmetry. The finite temperature correlators are described 
to order $T^2$ by the following mixing of zero-temperature correlators \cite{DE90}:
\begin{eqnarray} 
\Pi^{\mu\nu}_V (q; T)&=&(1-\epsilon)\,\Pi_V^{\mu\nu} (q; T=0)
\,+\,\epsilon\,\Pi_A^{\mu\nu} (q; T=0) \label{MIXV}\\
\Pi^{\mu\nu}_A(q; T)&=&(1-\epsilon)\,\Pi_A^{\mu\nu} (q; T=0)
\,+\,\epsilon\,\Pi_V^{\mu\nu} (q; T=0)\label{MIXA}
\end{eqnarray}
where $\epsilon=T^2/ 6 f^2_\pi$ is directly proportional to the scalar density
of the thermal pions. This implies that, to this order, the masses of the $\rho$ 
and $a_1$ meson do not change although the order parameters (quark condensate
and pion decay constant) are modified in contradiction with the BR scaling
law.  It is amusing to note that full mixing   $\epsilon=1/2$ 
corresponding to full symmetry restoration is realized at $T\simeq 160$ MeV 
very close to the lattice critical temperature. The above result has been extended 
beyond chiral limit in the chiral reduction formalism \cite{ZA96} in which the DPR writes~:  
\begin{eqnarray}
{dR\over d^4x d^4q}=& &-{\alpha^2\over \pi^3 \,q^2}\,{1\over e^{\beta q_0}+1}
\bigg[Im\,\Pi(q^2)\,-\, {2\over  f^2_\pi} \,\int {d{\bf k}\over (2\pi)^3}\, 
{n(\omega_k)\over\omega_k}\,Im\,\Pi_V(q^2)\nonumber\\
& &+\,{1\over \, f^2_\pi} \,\int {d{\bf k}\over (2\pi)^3}\, {n(\omega_k)
\over\omega_k}\,\left(Im\,\Pi_A((q+k)^2)\,+\,
Im\,\Pi_A((q-k)^2)\right)\,+....\bigg].
\end{eqnarray} 
The first term corresponds to the full electromagnetic correlator (with $\rho,  
\omega$ and $\phi$ pieces) and the  second term exhibits the quenching of
the (isovector-)vector correlator.  The last term represents the axial-vector
mixing  beyond the soft pion limit ($k \to 0$). The integration over all the
pion momenta yields a broadening of the (rho meson) spectrum which has to be
understood as an unavoidable consequence of partial chiral symmetry restoration.

\medskip
\noindent
{\it QCD sum rule}. 
The QCD sum rule approach aims at an understanding of
physical current-current correlation functions in terms of QCD by relating the
observed (or calculated) hadron spectrum to fundamental condensates $C_n$ 
(quarks, gluons) {\it i.e.} to the non-perturbative QCD vacuum structure. For large
space-like momenta ($Q^2=-q^2$), OPE techniques lead to~:  
\begin{eqnarray}
{\Pi(q^2=-Q^2)\over Q^2}&=&
\int_0^\infty\,{ds\over s}\,
{\left(-{1\over\pi}\right)\, Im\,\Pi(s)\over s\,+\,Q^2}
\nonumber\\
&=&{d_V\over 12\, \pi^2}\left[-C_0\,ln\left({Q^2\over \mu^2}\right)
\,+\,{C_1\over Q^2}\,+\,{C_2\over Q^4}\,+\,{C_3\over Q^6}\,+.....\right]
\label{SUMR}
\end{eqnarray}
I will not discuss here the technical difficulties and limitations 
of the method which can be applied  both in the vacuum and in the medium.
I will only emphasize that it provides a crucial test of consistency
between hadronic correlators and fundamental properties such as chiral 
symmetry and its restoration. Although such a test should be systematically done,
the approach is only of little predictive power. It has been found \cite{KL97,LEUP} 
that the
generic decrease of the quark and gluon condensates on the right-hand-side 
is compatible with the phenomenologically left-hand-side if either (a) the vector
meson masses decrease (together with small resonance widths as in the $\omega$
meson case) or (b) both width and
mass increase (as found in most phenomenological models for the $\rho$ meson).

\medskip
\noindent
{\it Dropping mass scenario.} 
Early QCD sum rule analysis based on a sharp ansatz for the vector mesons \cite{HA92}
gave a decrease of the vector meson (rho and omega) of about $20\%$ at normal 
nuclear matter density. At this time this result has been understood as 
 being in favor of the scaling law proposed by Brown and Rho \cite{SCAL} on the basis of
broken scale invariance in QCD~: 
\begin{equation}
\,{f^*_\pi\over f_\pi}\,=\,  
{m^*_\sigma\over m_\sigma}\,=\,{m^*_\omega\over m_\omega}
\,=\,{m_\rho^*\over m_\rho}
\,=\,\left({<\bar q q>^*\over <\bar q q>}\right)^{1/3}\,. 
\end{equation} 
The rho mass itself plays the role of an order parameter. Although these scaling 
laws contradict some low density results, chiral symmetry does not forbid the
vanishing of the mass at full restoration. But, as already discussed, such a
dropping mass scenario is certainly not a necessary consequence of chiral
symmetry restoration. Although this scenario yields a reasonable agreement 
with data \cite{PREM,BR97}, other hadronic many-body approaches containing other aspects of 
chiral symmetry 
restoration without dropping mass are also able to reproduce the data.
 \noindent
\begin{figure}
    \centering\epsfig{figure=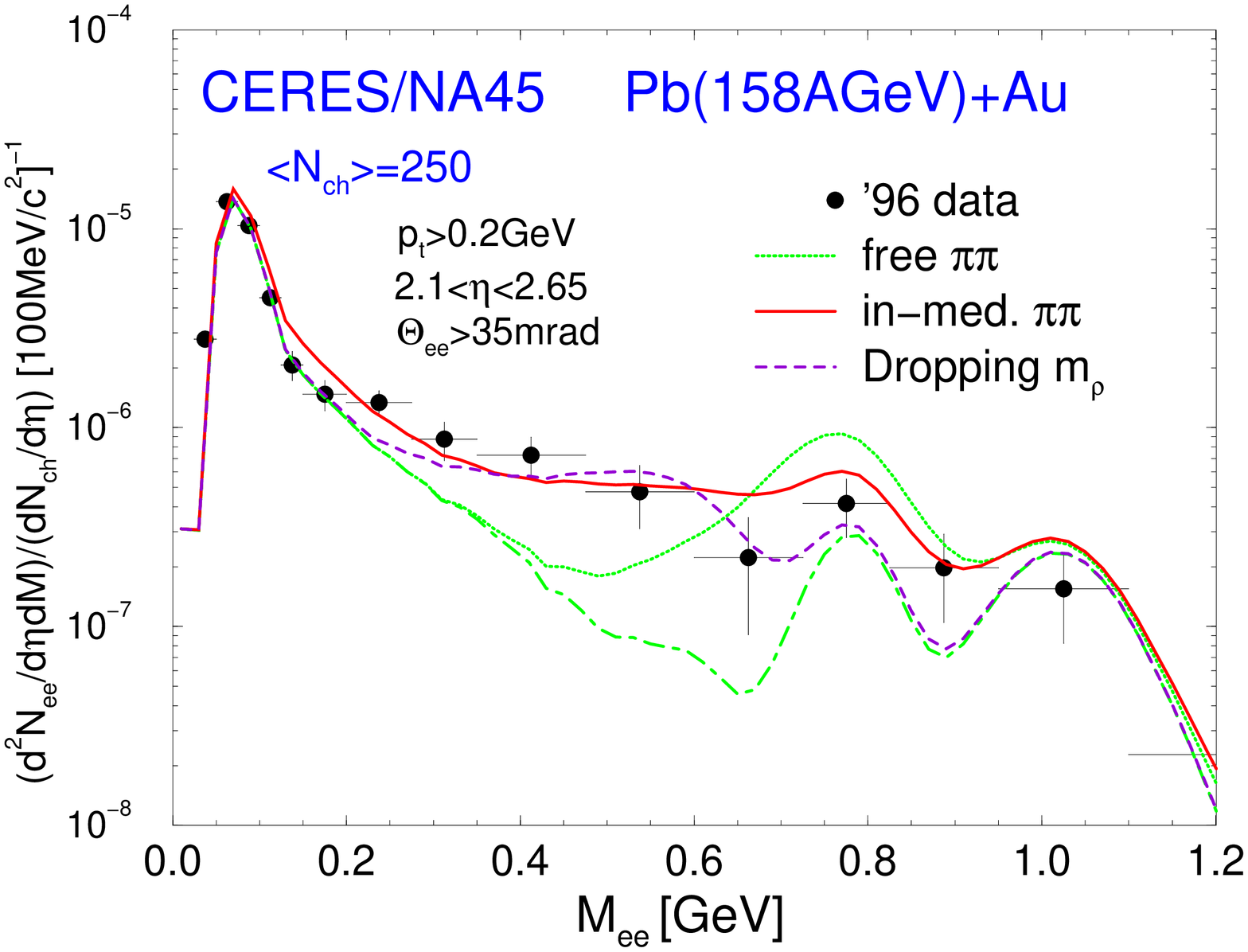,width=9cm, angle=270}
    \centering\epsfig{figure=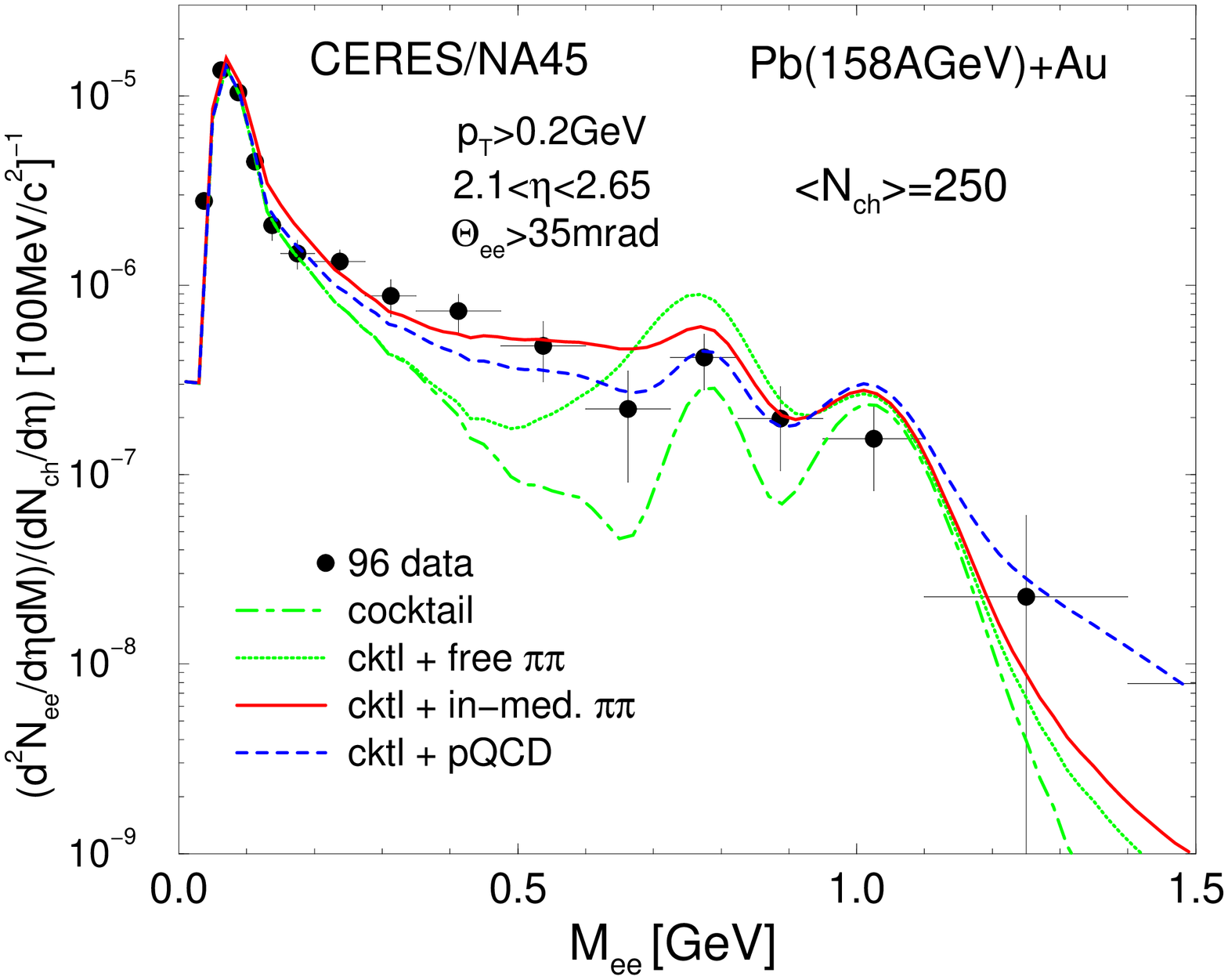,width=9cm,angle=270}
  \caption{Dilepton invariant mass spectrum measured in central 158 AGeV Pb+Au
  collisions compared to theoretical calculations \cite{RAWA}. The dashed-dotted
  lines
  correspond to the hadronic cocktail (without $\rho$ decays),  the dashed lines
  to the hadronic cocktail plus free $\pi\pi$ annihilation
  and the solid lines incorporate medium effects in the rho meson propagator
  (many-body approach). 
  In the upper panel the long-dashed line is obtained within the dropping
  mass scenario and in the lower panel the long-dashed line is an exploratory
  calculation using lowest order $q \bar q\to ee$ annihilation rates only within  
  the same expanding thermal fireball.}
       
\end{figure}  
      
\medskip
\noindent
{\it Many-body approaches}. More conservative approaches reside on standard many-body 
techniques to calculate the self-energy and consequently the rho meson spectral function. 
 They are based on effective hadronic Lagrangians possessing chiral symmetry and
incorporating vector dominance. The various parameters (coupling constants and form factor
cutoffs) are constrained as much as possible by other data and phenomenology 
(decay rates, photoabsorption,
$\rho N$ scattering \cite{FRI00},...). One is thus able to evaluate the in-medium 
 modification of the 
rho meson from its coupling to the various many-body excitations of the dense and hot
matter from which one gets the DPR at a given temperature and baryonic chemical potential.
As we will see below, once various final state hadronic decays are incorporated on top of 
the interacting fireball contribution,  
such an approach is able to reproduce the enhancement of the DPR below the 
$\rho/\omega$  peak and the apparent depletion of the peak itself. The latter depletion 
has a pure
many-body origin~: the propagator formalism generates resummations to all orders which are
totally absent in any kind of low-density expansion and/or in incoherent summations of
various processes.  
The total thermal yield in heavy-ion reactions is obtained by a space-time integration over
the density-temperature profile for a given collision system modeled within 
transport or hydrodynamics simulations. Another very successfull attempt is provided 
by a simple expanding thermal fireball \cite{CRW2} allowing to incorporate 
in a rather simple way 
hadronic many-body effects  which are needed to obtain a consistent description of the
data. In the most recent calculation \cite{RAWA} the trajectory 
starts at $(T,\rho_B)_{in}$=(190 MeV,
2.55$\rho_0$), goes through the experimentally deduced point in hadro-chemical analysis 
\cite{STAC}
up to thermal freeze out $(T,\rho_B)_{fo}$=(115 MeV, 0.33$\rho_0$).  Transport calculations, 
where no assumption is made about the degree of thermalization, 
also get better agreement with data
once in-medium spectral function is incorporated \cite{CBRW}.
It is convenient at least qualitatively to separate temperature  and baryonic density 
effects.

In a hot meson gas  the first
medium effect is the Bose enhancement of the $\pi\pi$ annihilation or, 
in terms of the rho
meson propagator, the temperature effect affecting the two-pion loop
contribution to the rho  self-energy. In addition $\rho$-meson scattering, in
first rank $\rho \pi\to \omega, a_1$, also significantly contributes to the rho
self-energy. It is important to notice that the $a_1$ piece is of   axial-vector
mixing  nature and the corresponding contribution to the DPR can be  seen as a consequence
of partial chiral symmetry restoration. 

Historically the first advocated baryonic density effect 
was the medium
modification of the two-pion loop through p-wave coupling of the pion to $\Delta$-h 
states. This effect, usually referred as the pion cloud contribution, gives a
significant enhancement of the DPR below the rho peak \cite{CHAN} and is mainly related to the
coupling of the rho to dressed  pions and $\Delta$-h states (the so-called pisobars). 
Again, the very important
point is that it contains an explicit mixing between the vector and the axial baryonic
current, as recently demonstrated \cite{CH99}. Finally, the direct coupling of the 
rho to baryonic resonances having sizable coupling to the rho
has been incorporated. In a many-body language the rho couples to $N^*-h$ excitations
building up the so-called rhosobars \cite{PIRN}. Among them the $N^*(1520)$ plays a prominent role
(since the coupling is of s-wave nature) and gives a very important contribution to the 
low mass enhancement \cite{PE98}. Contrary to the case of the $a_1$ and pion cloud contributions, 
the
connection with chiral symmetry restoration of the $N^*(1520)$ is not transparent. 
The resulting full spectrum which incorporates all the above effects within the
expanding fireball model  nicely accounts for the data (fig.7). 
A very important observation is that
this spectrum  
is very flat and very close to a pure perturbative quark-gluon spectrum (see lower panel of fig.7). 
One possible conclusion 
is that chiral symmetry restoration manifests itself 
as a lowering of the quark hadron duality
threshold from its free space value of 1.5 GeV  down to 0.5 GeV near the phase
boundary \cite{RAWA}.

\section{Conclusion}
We have seen the prominent role of in-medium chiral dynamics to generate 
strong reshaping of
hadronic spectral functions. Chiral symmetry restoration itself  yields a softening and
a sharpening of the scalar-isoscalar modes and the structure seen in ($\pi,\pi\pi$)
has been tentatively attributed to precursor effects of the restoration associated to
strong fluctuations of the chiral condensate. However, most of the time one observes a
broadening and a flattening of hadronic spectral functions. This is in particular 
true in the rho meson channel and convincing arguments based on detailed calculations
yield to the conclusion that dilepton spectra in relativistic heavy-ion collisions
probing the phase transition region constitute a possible signature of this
restoration associated to a lowering of the quark-hadron duality threshold.  
Nevertheless it is clear that a strong effort has to be pursued to improve theoretical analysis in
connection with present and forthcoming experimental data. For instance, it is not clear that the
($\pi^-,\pi^0 \pi^0$) data
 \cite {NI99} can be described within the  theoretical framework of section 3.
For what concerns dilepton production the connection of the resonance contribution 
(especially the $N^*(1520)$) with chiral symmetry restoration and axial-vector mixing 
remains to be fully elucidated although some suggestions have been proposed
 \cite{KI00}. In addition, on a more practical side,
the precise contribution of the omega meson should be
separated to isolate the interesting medium effects relative to the rho meson channel. 
In that respect forthcoming high resolution measurements in a more baryon-dominated regime 
will certainly bring  crucial information. In particular dilepton data obtained with the HADES
detector at GSI will be of utmost importance for studying  in the most favorable regime  
the baryonic medium effects which 
 already seem to play a dominant role in the CERN/SPS regime. 
  
\medskip
\noindent
Acknowledgments. I am especially grateful to D. Davesne and M. Ericson for critical comments 
on the manuscript.
I have also benefited from productive conversations  with P. Schuck, J. Wambach and J.
Delorme.

\end{document}